%Paper: astro-ph/9407026
%From: GASPERINI@to.infn.it
%Date: Fri, 8 Jul 1994 19:11:43 +0200 (WET-DST)

\magnification=1200
\hsize 15true cm \hoffset=0.5true cm
\vsize 23true cm
\baselineskip=15pt

\font\small=cmr8 scaled \magstep0
\font\grande=cmr10 scaled \magstep4
\font\medio=cmr10 scaled \magstep2
\outer\def\beginsection#1\par{\medbreak\bigskip
      \message{#1}\leftline{\bf#1}\nobreak\medskip\vskip-\parskip
      \noindent}

\def \me {\buildrel <\over \sim}
\def \Me {\buildrel >\over \sim}

\def \ra {\rightarrow}

\def \pr {\prime}

\def \ap {\alpha^\prime}

\def \ga {\gamma}

\def \r {\rho}

\def \noi {\noindent}

\def\sqr#1#2{{\vcenter{\hrule height.#2pt\hbox{\vrule width.#2pt
height#1pt \kern#1pt\vrule width.#2pt}\hrule height.#2pt}}}

\def\lsim{\mathrel{\rlap{\lower4pt\hbox{\hskip1pt$\sim$}}
    \raise1pt\hbox{$<$}}}         %less than or approx. symbol
\def\gsim{\mathrel{\rlap{\lower4pt\hbox{\hskip1pt$\sim$}}
    \raise1pt\hbox{$>$}}}         %greater than or approx. symbol

\nopagenumbers

\line{\hfil  DFTT-29/94}
\line{\hfil July 1994}
\vskip 2 cm
\centerline {\grande  The Inflationary Role of the Dilaton}
\vskip 0.5 true cm
\centerline{\grande in String Cosmology}

\vskip 1.5true cm
\centerline{M.Gasperini}
\centerline{\it Dipartimento di Fisica Teorica, Universit\'a di Torino,}
\centerline{\it Via P.Giuria 1, 10125 Turin, Italy,}
\centerline{\it and}
\centerline{\it INFN, Sezione di Torino, Turin, Italy}

\vskip 1.5 true cm
\centerline{\medio Abstract}

\noindent
It is stressed that the kinematical problems of the standard
cosmological model can be solved by a phase of accelerated
contraction of the cosmic scale factor. Such a behaviour is only a
particular case of a more general ``pre-big-bang" inflationary scenario,
arising naturally in a string cosmology context, and driven
asymptotically by the free dilaton field without any self-interaction
potential and/or vacuum contribution.

\vskip 2 true cm
\centerline{----------------------------------}
\vskip 1 true cm
\noi
To appear in the {\bf Proceedings of the First Int.
Workshop on Birth of the Universe and Fundamental Physics},
{\it Roma, May 18-21, 1994} (Springer-Verlag, Berlin)

\vfill\eject

\footline={\hss\rm\folio\hss}
\pageno=1

\centerline{\bf THE INFLATIONARY ROLE OF THE DILATON IN STRING COSMOLOGY}
\bigskip
\centerline{M. GASPERINI}
\centerline{\small{\it Dipartimento di Fisica Teorica, Universit\'a di
Torino, }}
\centerline{\small{\it Via P.Giuria 1, 10125 Turin, Italy,}}
\centerline{\small{\it and INFN, Sezione di Torino, Turin, Italy}}
\bigskip
\centerline{ABSTRACT}
\midinsert
\narrower
\noi
It is stressed that the kinematical problems of the standard
cosmological model can be solved by a phase of accelerated
contraction of the cosmic scale factor. Such a behaviour is only a
particular case of a more general ``pre-big-bang" inflationary scenario,
arising naturally in a string cosmology context, and driven
asymptotically by the free dilaton field without any self-interaction
potential and/or vacuum contribution.
\endinsert
\bigskip
\noi
{\bf 1. Introduction: Inflation Without Inflation}

\noi
It is well known that various problems of the standard cosmological
scenario can find a natural explanation [1], if we assume that the phase
of radiation dominance is preceeded in time by a phase of accelerated
expansion (the so-called inflation), characterized by $H>0$,
$\ddot a>0$, where $a$ is the scale factor of an isotropic metric, $H=
\dot a /a$, and the dot denotes differentiation with respect to cosmic
time. What seems to be presently less known, however, is the existence
of another possible solution of the standard kinematical problems
(suggested by recent studies of string cosmology [2,3,4]), based on the
simple assumption that the evolution of our universe is time-symmetric
with respect to some given past curvature scale $H_1\sim 1/t_1>0$.

Suppose, in particular, that our universe is radiation-dominated up to
$H_1$, i.e. $a_+(t)\sim (t+t_1)^{1/2}$ for $t>0$. By time symmetry, the
scale factor is to be continued to negative times as
$$
a_-(t)\sim (t_1-t)^{1/2}, ~~~~~~~~~~~~~~~t<0 \eqno(1.1)
$$
This metric describes accelerated contraction, $\dot a<0$, $\ddot a<0$,
and growing curvature scale, $|\dot H|>0$, up to $H_1$, and it is
still an exact
solution of the radiation-dominated cosmological equations (with
negligible spatial curvature), as the Einstein equations are invariant
under time-reversal transformations. Of course, near $t=0$, a suitable
modification of both $a_{\pm}$ and of the source stress tensor is to be
expected, in order to guarantee the presence of a bounce at $|H|\sim H_1
$, and a continuous matching of the two branches of the radiation-
dominated evolution. As a consequence, $a_{\pm}$ are to be regarded only
as the asymptotic form of the solution, for $|H|<<H_1$. This
asymptotic form is already enough, however, to show that the horizon and
flatness problems [1] are absent in a radiation-dominated universe, if
its evolution is time-symmetric.

In fact, in the primordial contracting phase (1.1) the proper size of a
causally connected region shrinks in time like the scale factor, while
the proper size $d_e(t)$ of the event horizon,
$$
d_e(t)=a(t)\int_t^{t_1}{dt^\pr \over a(t^\pr)}={1\over 2}(t_1-t)\eqno
(1.2)
$$
shrinks linearly in cosmic time, i.e. faster than $a(t)$. This means
that causally connected scales are ``pushed out" of the horizon (just
like in conventional inflationary expansion). The horizon problem may
thus be solved, provided the ratio
$$
{a(t)\over d_e(t)}\sim \eta^{-1}\eqno(1.3)
$$
($\eta$ is the conformal time coordinate, defined by $dt=ad\eta$) grows
enough during the contraction ($t<0$), so as to compensate the
subsequent decrease of $a(t)/t$ during the phase of decelerated
expansion ($t>0$). To obtain, in particular, a causally
connected universe at the end ($\eta=\eta_2$) of the radiation-dominated
epoch, we must require according to (1.3) that the contracting phase is
at least as long as the subsequent expanding one, namely
$$
{|\eta_i|\over |\eta_1|}
\Me {|\eta_2|\over |\eta_1|}
\eqno(1.4)
$$
where $\eta_i$ marks the beginning of the contraction, and $\eta_1$ is
the fixed point of the time-reversal symmetry, corresponding to the
maximum scale $H_1$. This condition is certainly satisfied if the
evolution is time-symmetric, irrespective of the value of $\eta_1$.

With similar arguments one can show that also the flatness problem is
solved by a time-symmetric cosmological evolution. Indeed, the solution
of this problem requires that the ratio between the spatial curvature
term, $k/a^2$, and the other terms of the cosmological equations,
$$
{k\over a^2H^2}={k\over \dot a^2}\sim \eta^2 \eqno(1.5)
$$
at the end of the contracting phase ($\eta=\eta_1$) is tuned to a value
small enough, so that the subsequent decelerated evolution eventually
leads to a value of the ratio $\me 1$ at $\eta=\eta_2$. This implies
$$
\left(\eta_1\over \eta_i\right)^2
\me \left(\eta_1\over \eta_2\right)^2
\eqno(1.6)
$$
which is equivalent to the condition (1.4), and which is again satisfied
by a time-symmetric cosmological evolution.

The presence of a long enough period of accelerated contraction, which
may follow in particular from the hypothesis of time-symmetric cosmology
(like in the previous example), has thus all the kinematical effects of
inflation, with no need of introducing some ``ad hoc" inflaton
potential, slow-rolling conditions, fine-tuning, and so on, like in
conventional inflationary models [5]. A possible objection to the
complete equivalence of inflation and time-reversal symmetry is that the
effective potential, $V(\eta)=(d^2a/d\eta^2)/a$, appearing in the scalar
and tensor perturbation equations [6], is vanishing both for the
expanding and contracting radiation-dominated evolution, in which $a\sim
|\eta|$. This implies that no parametric amplification of the
perturbations is possible, in such context. However, in a truly time-
symmetric evolution the radiation-dominated contraction should be
preceeded by the time-reversed of the dust-dominated evolution, which also
describes accelerated contraction with $a\sim |t|^{2/3}\sim |\eta|^2$.
The transition between the two contracting phases amplifies
perturbations with just a flat Harrison-Zeldovich spectrum, even if only
for those comoving scales $k$ which are smaller than the ``height"
$|\eta_2|^{-1}$ of the effective potential barrier, namely for scales
entering inside the horizon during our present matter-dominated epoch.
\vskip 1 cm
\noi
{\bf 2. The Pre-Big-Bang Scenario}

\noi
Another objection to the ``inflationary" properties of a phase of
accelerated contraction is that such a phase seems unable to dilute the
possible production of topological defects, such as monopoles or
strings. A possible answer to this objection is that strings shrinks in
contracting backgrounds and, as a consequence, they are eventually
diluted by the same amount as in the conformally related
superinflationary (expanding) backgrounds [3]. Moreover, if it is true
that monopoles
inflate together with the metric background in which they are embedded,
as recently argued in [7], then they should shrink, just like strings,
if the background is contracting. So accelerated contraction is not
incompatible, in principle, with the dilution of unwanted topological
relics.

However, what it should be stressed at this point is that the previous
example of phase with accelerated contraction and growing curvature,
obtained through a time-reversal transformation from the standard
radiation-dominated cosmology, is only a particular case of
``pre-big-bang" scenario [8,2-4]. Such a scenario, describing a universe
which starts from a flat and vacuum initial state, and approaches a
maximum curvature scale (identified with the big-bang) through a period
of accelerated evolution, finds a natural motivation in the context of
the equations obtained from the low energy string effective action
(rather than in the context of Einstein's cosmological equations). In
their minimal version such equations contain, besides the metric and the
matter sources, also a non-minimally coupled scalar field (the dilaton).
For a homogeneous and isotropic, spatially flat, $(d+1)$-dimensional
background, with perfect fluid sources ($p/\rho=\ga=const$), such
equations can be written explicitly as [2,8]
$$
(\dot \phi -dH)^2-dH^2=8\pi G \r e^\phi
$$
$$
2\dot H-2H(\dot \phi-dH)=8\pi G \ga \r e^\phi
$$
$$
(\dot \phi -dH)^2-2\ddot \phi +2d \dot H + dH^2=0 \eqno(2.1)
$$
(where $G$ is the $d$-dimensional analog of the Newton constant). They
are invariant not only under the usual time-reversal transformation
$t\ra -t$, but also under the ``duality" transformation [9,10]
$$
a \ra a^{-1},~~~~~~ \phi \ra \phi -2d\ln a,~~~~~ \ga \ra -\ga \eqno(2.2)
$$

By combining duality and time reversal transformations it is thus
possible, in a string cosmology context, to associate to any given
standard ``post-big-bang" solution, with decreasing curvature scale,
a related ``pre-big-bang" solution whose curvature has a specular
behaviour with respect to the chosen scale of time-symmetry, but which
is expanding, in an accelerated (inflationary) way. Consider, for
instance, the general solution of eqs.(2.1), which becomes, at low
enough curvature scales [2,3],
$$
a(t)\sim |t|^{2\ga /(1+d\ga^2)},~~~~ \phi
\sim {d\ga -1\over \ga}\ln a, ~~~~ \r\sim a^{-d(\ga+1)}\eqno(2.3)
$$
For $\ga>0$ and $t>0$ the metric describes a decelerated expansion with
decreasing curvature, characterized by
$$
\dot a >0,~~~~ \ddot a <0,~~~~ \dot H<0 \eqno(2.4)
$$
By performing the duality transformation (2.2), plus a time inversion,
we are led to a metric describing accelerated expansion
(of the so-called superinflationary type) and increasing curvature scale,
characterized by
$$
\dot a >0,~~~~ \ddot a >0,~~~~ \dot H>0 \eqno(2.5)
$$
The particular case of the radiation-dominated solution ($\ga=1/d$)
corresponds to a constant dilaton,
$$
a\sim t^{2/(d+1)},~~~~ \phi = const, ~~~~ \r \sim a^{-(d+1)},
{}~~~~ t>0 \eqno(2.6)
$$
and it is associated to a pre-big-bang phase dominated by a gas of
stretched strings [11] (with $\ga=-1/d$), characterized by
superinflationary expansion, growing curvature and dilaton coupling,
$$
a\sim (-t)^{-2/(d+1)},~~~~ \phi = 2 d \ln a, ~~~~ \r \sim a^{-(d-1)},
{}~~~~ t<0
\eqno(2.7)
$$

The particular case of pre-big-bang discussed in the previous section
($a\sim (-t)^{1/2}$, $\phi =const$) is thus only a trivial example
obtained from the solution (2.6) through time-reversal, which is the
only possible transformation in the context of the Einstein cosmological
equations, where there is no dilaton and the duality symmetry cannot be
implemented. In the string cosmology context, on the contrary, the dilaton
is present, and the associated duality symmetry has various important
consequences.

First of all, with a superinflating (instead of contracting)
pre-big-bang scenario the scale factor evolves monotonically
in time, and
this leads to the interesting possibility of self-dual cosmological
solutions [8], characterized by $a(-t)=a^{-1}(t)$, and connecting in a
smooth way the two duality-related pre and post-big-bang regimes (see
however [12] for non-monotonic but smooth solutions). Of course, a
superinflating pre-big-bang metric may become contracting when
transformed in the Einstein frame (where the dilaton is minimally
coupled to the metric). In that frame, however, the contraction is
accelerated and the curvature is still growing, so that the solution of
the standard kinematical problems is preserved [2-4], as discussed also
in Section 1.

The presence of a non-trivial dilaton background, moreover, contributes
to the effective potential appearing in the linearized perturbation
equation [13,2,3], and then to the parametric amplification of the
perturbations. The potential corresponding to the solution (2.7), in
particular, is no longer flat, unlike the case of the solution (2.6)
(for $d=3$). But the main consequence of a dilaton background is that,
in the high curvature regime, the general solution of the
equations (2.1)
becomes dilaton-dominated (quite independently of the matter equation of
state), and the universe enters a phase of dilaton-driven
superinflation, with [2,3]
$$
a\sim (-t)^{-1/\sqrt d},~~~~~~~~~~~~~~~ \phi \sim (d+\sqrt d)\ln a
\eqno(2.8)
$$
When the sources are, in particular, a diluted gas of non-interacting
fundamental strings, and we impose as initial condition the flat and
cold string perturbative vacuum ($a=const$, $\phi= -\infty$), then the
general solution of the system of background string equations and string
equations of motion shows that the beginning of the superinflationary
evolution of the metric, and of the phase of dilaton dominance, are
nearly coincident at the same time scale [3]. In that case the dilaton
field is completely responsible for the phase of accelerated evolution.
\vskip 1 cm
\noi
{\bf 3. Discussion}

\noi
According to the string cosmology equations, a phase of dilaton-driven
isotropic superinflation can thus be arranged with no need of a
simultaneous shrinking of the internal dimensions (like in the
conventional realizations of superinflation [14]) or of some fine-tuned
dilaton potential. Indeed, at low energy scales (with respect to the
string tension $(\ap)^{-1/2}\simeq 10^{-1}~M_p$) the dilaton potential
can appear at a non-perturbative level only, and it is expected to be
negligible, as $V(\phi)\sim \exp [-\exp(-\phi)]$.

A string-motivated dilaton potential must play a fundamental role,
however, in the transition
from the accelerated evolution with growing curvature
of the pre-big-bang regime, to the
subsequent standard cosmological evolution. As recently discussed
in [15], a smooth transition seems to be impossible without including
in the low energy
string effective action  higher curvature corrections
and the contribution of a non-perturbative dilaton potential. The
``graceful exit" from the pre-big-bang to the post-big-bang regime is
indeed the main unsolved problem, at present, of this scenario, even if
recent results on exact string solutions (to all orders in $\ap$) are
encouraging [16]. The non-perturbative potential, in particular, is
expected to force the dilaton to a minimum which gives it a mass, and
freezes the value of the Newton constant to its present value.

In addition, the transition from the growing to decreasing curvature
regime should give rise to a copious non-adiabatic production of
radiation, in order to explain the large amount of entropy presently
observed on a cosmological scale. An interesting possibility is that
such entropy be produced in the decay of the dilatons (or of other
massive particles) created from the vacuum by the expanding metric
background [3,17]. In such case, the dilaton freezing is likely to occur
not simultaneously to, but sensibly after the bounce of the curvature at
a maximum scale (of Planckian order); moreover, the universe is expected
to be dilaton-dominated even in the first period of the post-big-bang
era, before the radiation production. But for what concerns these
aspects of string cosmology work is still in progress [18], and I hope to
be able to report more details in some future publication.

\vskip 1 cm
\noi
{\bf Acknowledgements.}

\noi
I am very grateful to M. Giovannini and G. Veneziano for a
fruitful collaboration which led to the results reported
in this paper. I wish to thank also F. Occhionero for his kind
invitation, and for the perfect organization of this excitant
Workshop on the various aspects of primordial cosmology.

\vskip 1 cm
\centerline{\bf References.}

\item{1.}A. Guth, Phys. Rev. D23 (1981) 347

\item{2.}M. Gasperini and G. Veneziano, Mod. Phys. Lett. A8 (1993) 3701

\item{3.}M. Gasperini and G. Veneziano, Phys. Rev. D50 (1994) No.4

\item{4.}See also M. Gasperini, ``Phenomenological aspects of the
pre-big-bang scenario in string cosmology", to appear in Proc. of the
2nd Journ\'ee Cosmologie, Paris, June 1994 (World Scientific, Singapore)
(astro-ph/9406056)

\item{5.}See for instance E. W. Kolb and M. S. Turner, ``The early
universe" (Addison-Wesley, Redwood City, Ca, 1990)

\item{6.}See for instance V. Mukhanov, H. A. Feldman and R. Brandenberger,
Phys. Rep.
215 (1992) 203

\item{7.}A. Linde, ``Monopoles as big as the universe", Preprint
SU-ITP-94-2 (astro-ph/9402031);

A. Vilenkin, ``Topological inflation", CTP-MIT Preprint
(hep-th/9402085)

\item{8.}M. Gasperini and G. Veneziano, Astropart. Phys. 1 (1993) 317

\item{9.}G. Veneziano, Phys. Lett. B265 (1991) 287

\item{10.}M. Gasperini and G. Veneziano, Phys. Lett. B277 (1992) 256

\item{11.}M. Gasperini, N. S\'anchez and G. Veneziano, Int. J. Mod. Phys.
A6 (1991) 3853; Nucl. Phys. B364 (1991) 365

\item{12.}C. Angelantonj, these Proceedings. See also Diploma Thesis,
Universita' dell'Aquila (1994) (unpublished)

\item{13.}M. Gasperini and M. Giovannini, Phys. Rev. D47 (1993) 1529

\item{14.}D. Shadev, Phys. Lett. B317 (1994) 155;

R. B. Abbott, S. M. Barr and S. D. Ellis, Phys. Rev. D30 (1994) 720;

E. W. Kolb, D. Lindley and D. Seckel, Phys. Rev. D30 (1994) 1205;

\item{15.}R. Brustein and G. Veneziano, ``The graceful exit problem in
string cosmology", CERN-TH.7179/94

\item{16.}E. Kiritsis and C. Kounnas, ``Dynamical topology change in
string theory", CERN-TH.7219/94
(hep-th/9404092)

\item{17.}M. Gasperini, Phys. Lett. B327 (1994) 214

\item{18.}M. Gasperini, M. Giovannini and G. Veneziano, in preparation

\end